\documentclass[final,1p,times]{elsarticle} 
\usepackage{graphicx} 
\usepackage{amssymb} 
\usepackage{amsthm} 
\usepackage{subfigure}

\journal{Nuclear Physics A} 
\begin{document} 

\begin{frontmatter} 


\title{Effects of partial thermalization on HBT interferometry}


\author[a]{\underline{Cl\'ement Gombeaud}}
\author[a,b]{Tuomas Lappi}
\author[a]{Jean-Yves Ollitrault}

\address[a]{Institut de Physique Th\'eorique, CEA/DSM/IPhT,
  CNRS/MPPU/URA2306\\ CEA Saclay, F-91191 Gif-sur-Yvette Cedex.}
\address[b]{Department of Physics
 P.O. Box 35, 40014 University of Jyv\"askyl\"a, Finland}

\begin{abstract} 
Hydrodynamical models have generally failed to describe interferometry radii measured at
RHIC. In order to investigate this ``HBT puzzle'', we carry out a systematic study of HBT
radii in ultrarelativistic heavy-ion collisions within a two-dimensional transport model. We
compute the transverse radii $R_o$ and $R_s$ as functions of $p_t$ for various values of the Knudsen
number, which measures the degree of thermalization in the system.
For realistic values of the Knudsen number estimated from $v_2$ data, we obtain $R_o/R_s \simeq 1.2$, much closer to data than standard hydrodynamical
results. Femtoscopic observables vary little with the degree of thermalization. Azimuthal
oscillations of the radii in non central collisions do not provide a good probe of thermalization. 

\end{abstract} 

\end{frontmatter} 



\section{A short reminder on the model}

In this talk, we review our recent calculation~\cite{Gombeaud2009} of HBT radii, using a transport model described in more details in~\cite{Gombeaud:2007ub}.
We model the expansion of the matter created in the central rapidity region of a heavy-ion collision using a
  relativistic 2 dimensional Boltzmann equation with massless particles. 
The equation of state (EOS) of the medium is that of a conformal ideal gas $\varepsilon =2P$.
We use two different initial momentum distributions. The first is a locally thermal distribution 
in order to make comparison with hydrodynamic simulations~\cite{Gombeaud:2007ub}. The second
is a more realistic distribution based on a parametrisation of the  Color Glass Condensate (CGC) 
initial gluon spectrum from~\cite{Krasnitz:2001qu,Lappi:2003bi}.
In both cases, our momentum distribution is scaled such a way that the average 
transverse momentum per particle is given by $\langle p_t \rangle=420$MeV, which corresponds roughly to the value for pions at the top RHIC energy.

In our numerical solution of the 2+1 dimensional Boltzmann equation, the partonic cross section
is assumed to be constant and isotropic in the center of mass frame for simplicity.
The dynamics of the system is then controlled by a single quantity, the mean free path $\lambda$, 
which is the average distance travelled by a particle between two collisions.
The Knudsen number $K$ is the ratio of the mean free path to the characteristic size of the system.
Because $K^{-1}$ is proportional to the average number of collisions per 
particle during the evolution, we use $K$ to characterize the 
degree of thermalization of the system; hydrodynamics is the limit $K \rightarrow 0$ while the limit
$K \rightarrow \infty$ corresponds to the free-streaming regime.


For a particle with outgoing momentum ${\bf p}_t$, we denote by $(t,x,y)$ the
space-time  point where the last collision occurs. The ``out'' and
``side'' coordinates are then defined as the projections parallel and 
orthogonal to the particle momentum. The HBT radii are then obtained from rms widths of the space time distribution of particles 
with the same outgoing momentum. 
Radii defined in this way coincide with those obtained from the curvature of the correlation function at zero relative momentum.

\section{Central collisions}

In order to mimic a central Au-Au collision at RHIC, we use a
gaussian initial density profile with
$\sigma_x=\sigma_y=3$~fm. This value corresponds to the rms width of
the initial density profile in an optical Glauber calculation. 
\begin{figure}[!ht]
\centering
\subfigure{
\includegraphics[width=2.5in]{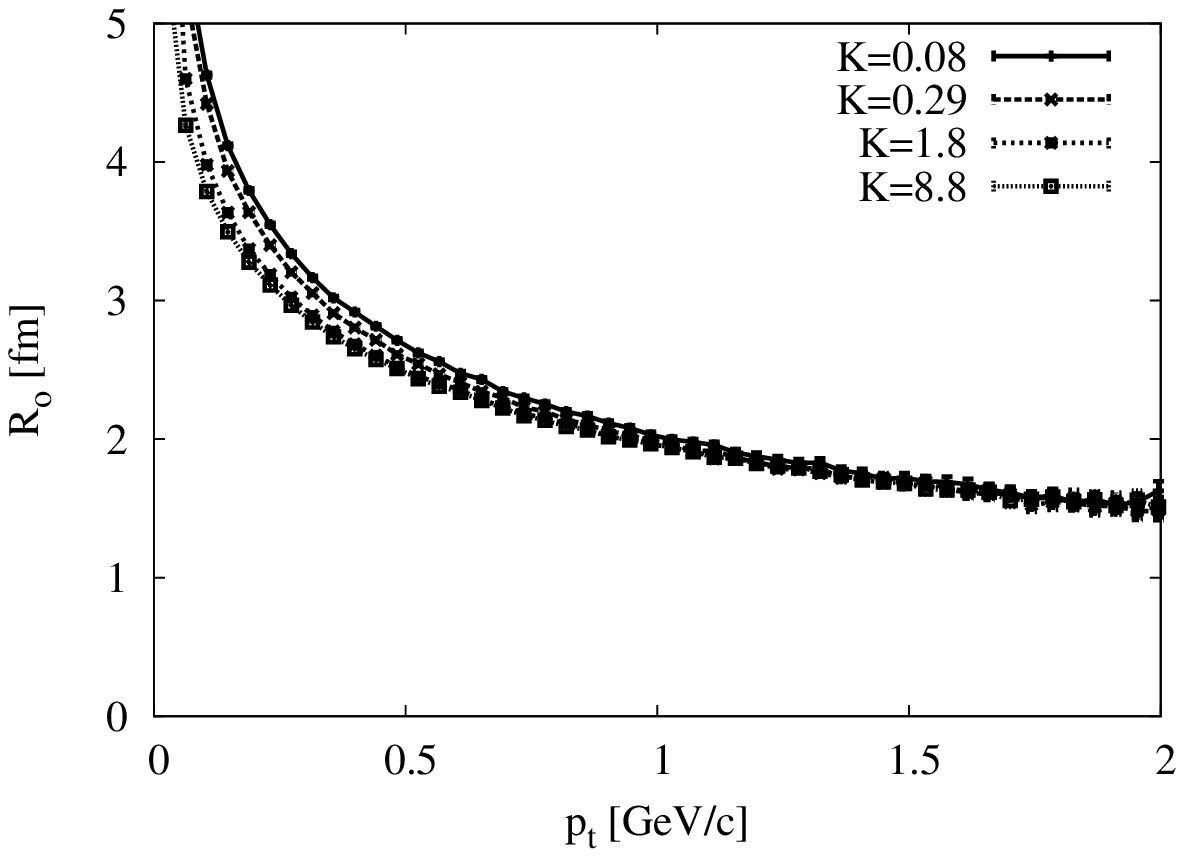}}
\subfigure{
\includegraphics[width=2.5in]{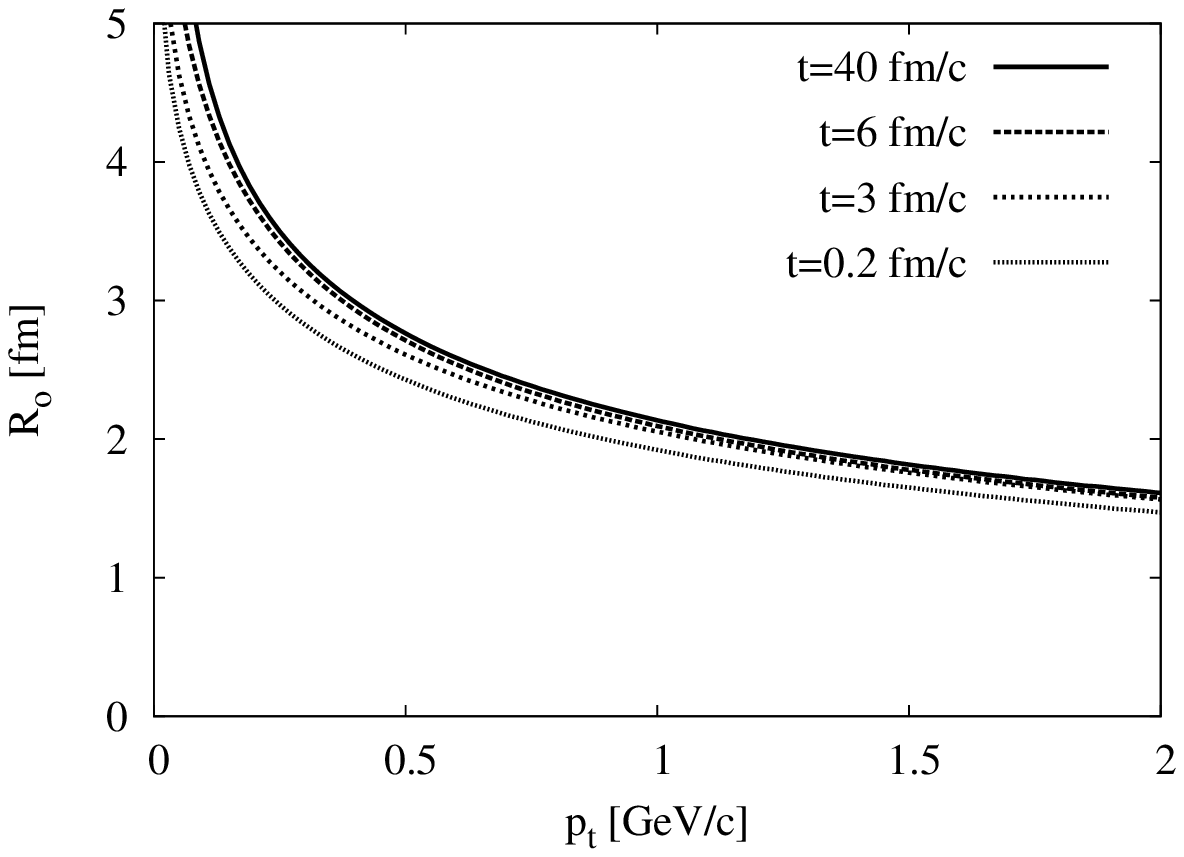}}
\caption{HBT radius $R_o$ versus  transverse momentum $p_t$ of
  particles: (left) Transport calculations; the curves are labeled by
  the value of the Knudsen number $K$. (right) Ideal hydrodynamics at fixed freeze out time $t$; the curves
  are labeled by the value of $t$.}
\label{fig:fig1}
\end{figure}

Figure~\ref{fig:fig1} (left) displays $R_o$ versus $p_t$ for different values of the Knudsen number.
It shows that $R_o$ decreases when increasing $K^{-1}$ or $p_t$.
The decrease with $p_t$ is usually considered as a signature of the hydrodynamical behaviour, but even if it 
is more pronounced for hydrodynamics, it is also seen in the free streaming regime.
Figure~\ref{fig:fig1} (right) displays $R_o$ versus $p_t$ for different values of the freeze-out time in an ideal hydrodynamic calculation.
Comparing figure~\ref{fig:fig1} (left) and (right), we see that hydrodynamics at $t=0.2\textrm{fm}/c$ gives the
same radii as transport for high values of $K$ and that hydrodynamics at $t=40\textrm{fm}/c$ gives almost the same results 
as transport calculations for small $K$. Our interpretation is that 
decreasing $K$ amounts to increasing the number of collisions per particles until the ``freeze out'', which 
has qualitatively the same effect as increasing the duration of the hydrodynamic phase.

\begin{figure}[!htb]
\centering
\subfigure{
\includegraphics[width=2.5in]{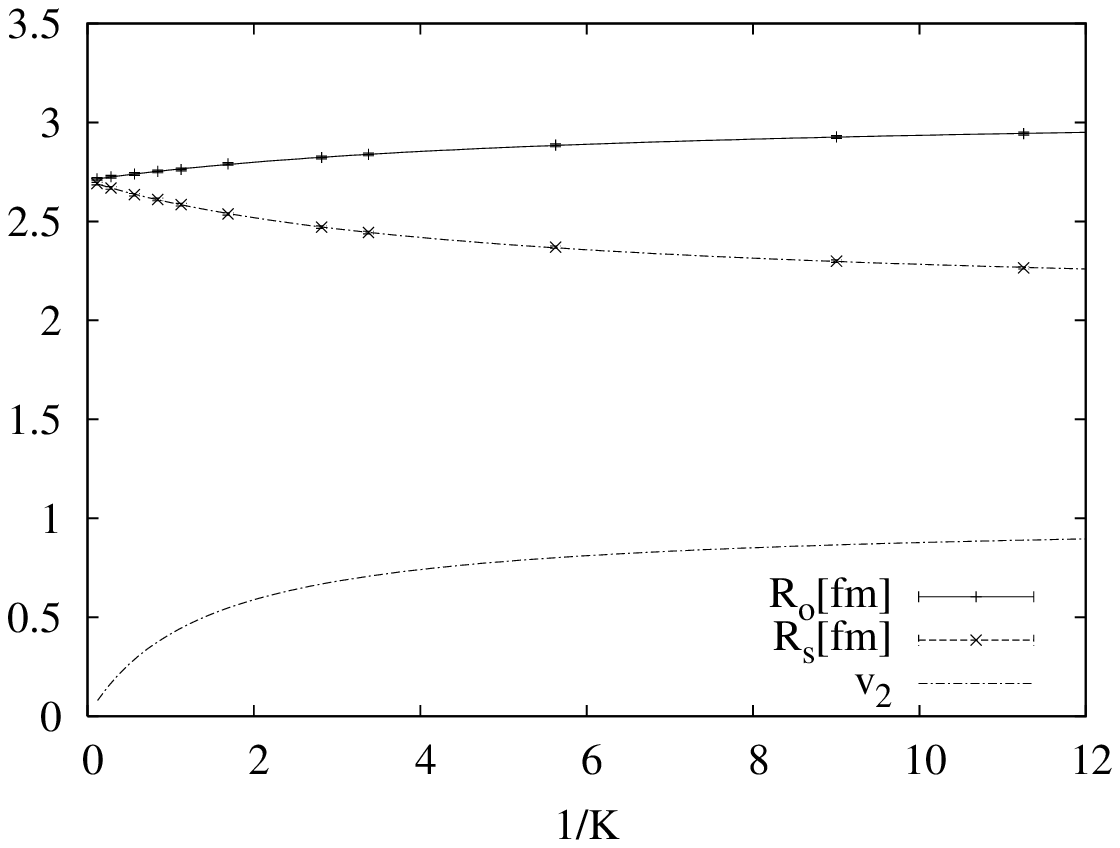}}
\subfigure{
\includegraphics[width=2.5in]{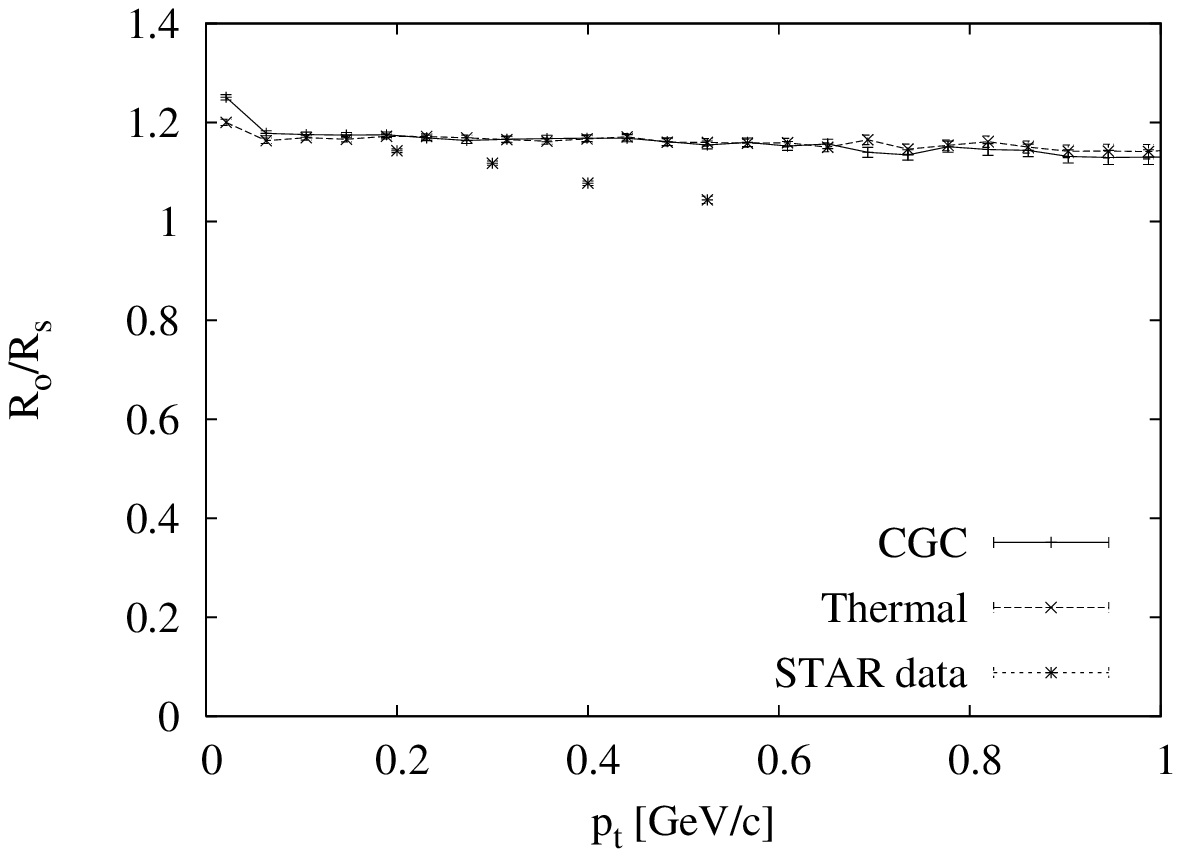}}
\caption{Results obtained for HBT radii averaged over the $p_t$ interval [0.25 ,0.75] GeV: 
(left) $R_o$ and $R_s$ versus the degree of thermalization. The dotted curve shows, for sake of
  illustration, the variation of elliptic flow in a non-central
  collision, scaled by the hydrodynamical limit
  (from~\cite{Gombeaud:2007ub}). 
(right) $R_o/R_s$ versus $p_t$ for $K=0.3$. Dashed line: thermal initial
  conditions; full line: CGC
  initial conditions. Stars: data
  from STAR~\cite{Adams:2004yc}}.
\label{fig:fig3}
\end{figure}

In the transport simulation, we observe that 
$R_o$ increases and $R_s$ decreases when the number of collisions per particle $1/K$ increases, as plotted on figure~\ref{fig:fig3} (left). 
On the same figure, we also plot the elliptic flow from~\cite{Gombeaud:2007ub} versus $K^{-1}$. 
One sees that $v_2$ saturates to the hydro limit much faster than $R_o$ and $R_s$. 
Hydrodynamical calculations~\cite{Hydro}~\cite{Hydro2}~\cite{Hydro3} usually yield a value of $R_o/R_s$ 
of the order of 1.5, while RHIC data are compatible with 1.
Figure~\ref{fig:fig3} (right) displays the HBT ratio $R_o/R_s$ with $K=0.3$, which is the value obtained by fitting the centrality dependence of elliptic flow~\cite{Drescher}. Our results are in much better agreement with 
experimental data than models based on ideal hydrodynamics. 
It has been recently argued~\cite{Broniowski:2008vp} that ideal hydrodynamics simulations with an early freeze out also explains the HBT puzzle.
Our results also show that increasing the Knudsen number amounts to decreasing the freeze out time. We find that partial thermalization,
which has been shown to explain the centrality dependence of $v_2$, also solves most of the HBT puzzle.
In an other recent work, Scott Pratt argued~\cite{Praat} that the HBT puzzle results from a combination of several factors. 
Our results show that partial thermalization alone solves most of the puzzle.

\section{Azimuthally sensitive HBT}

\begin{figure}[!htb]
\centering
\subfigure{
\includegraphics[width=2.5in]{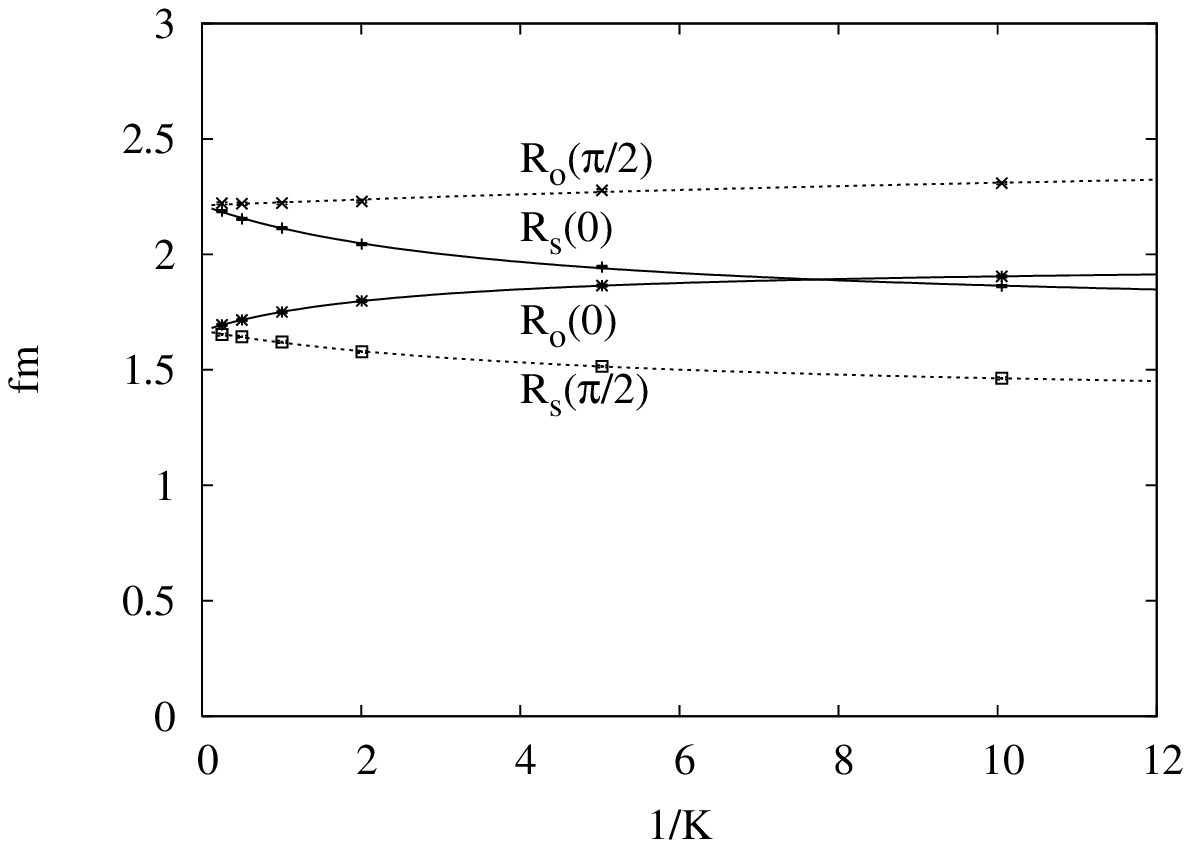}}
\subfigure{
\includegraphics[width=2.5in]{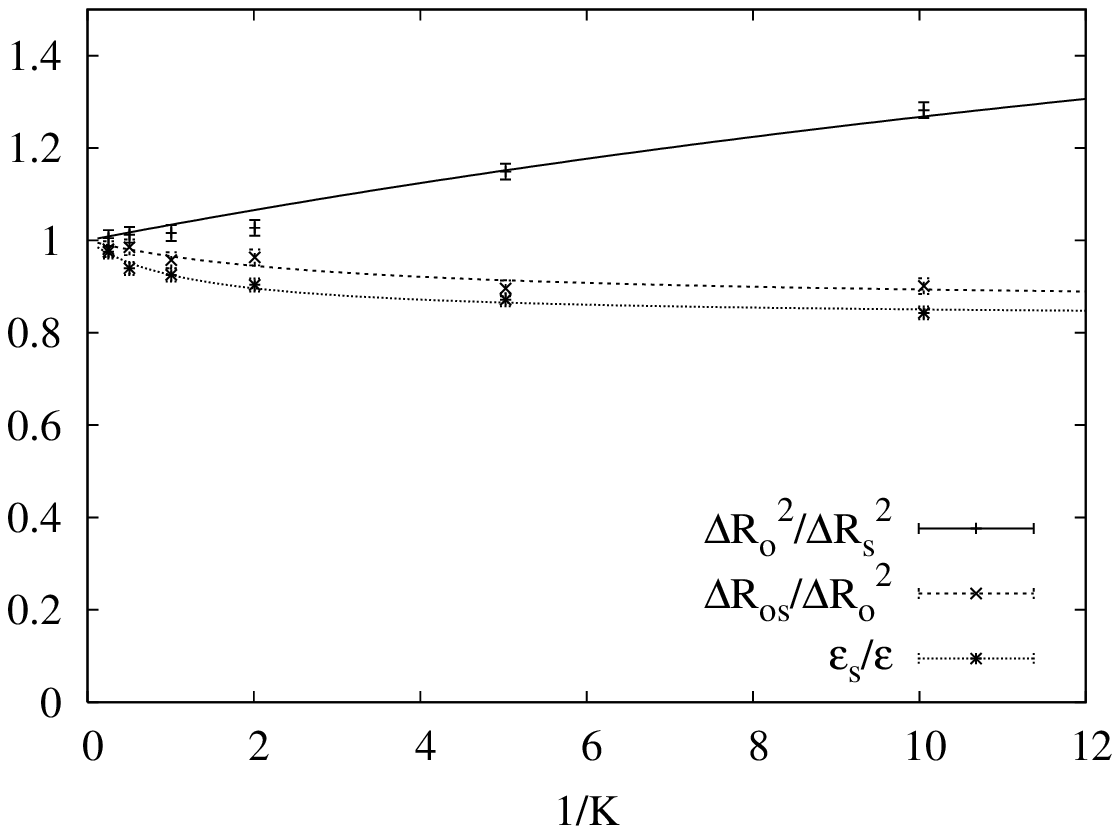}}
\caption{Evolution of AzHBT observables with the degree of thermalization:
(left) In-plane ($\phi=0$) and out-of-plane ($\phi=\pi/2$) radii
 for thermal initial conditions.
 (right) Ratios of oscillation amplitudes for thermal
  initial conditions.}
\label{fig:fig8}
\end{figure}

For a non-central collision, the interaction region is elliptic, and HBT radii depend on the azimuthal angle $\phi$, defined with respect to the impact parameter direction. 
Figure~\ref{fig:fig8} (left) shows the radii in plane and out of plane as functions of the average number of collisions per particle.
As in the case of central collisions, we observe that $R_o$ increases and $R_s$ decreases when increasing $1/K$, but the slope is 
larger for $\phi = 0$ than for $\phi =\pi /2$.
Our interpretation is that thermalization is faster in plane that out of plane due to flow effects.

We define three oscillations amplitudes:
\begin{eqnarray}
\label{oscamplitudes}
\Delta R_o^2 &=& R_o^2(\pi/2)-R_o^2(0)\cr
\Delta R_s^2 &=& R_s^2(0)-R_s^2(\pi/2)\cr
\Delta R_{os} &=& R_{os}(3\pi/4)-R_{os}(\pi/4).
\end{eqnarray}
These amplitudes scale with the eccentricity of the overlap area, which is not directly measured. We avoid this dependence
by taking ratios as shown on figure~\ref{fig:fig8} (right). These ratios are equal to unity in the free streaming regime and can also be 
measured experimentally. For realistic values of K in the range $[0.3,0.5]$, both ratios deviate little from unity. 
On figure~\ref{fig:fig8} (right) we also plot the ratio of the eccentricity seen in HBT radii
 $\epsilon_s=(R_{s}^2(0)-R_{s}^2(\pi/2))/(R_{s}^2(0)+R_{s}^2(\pi/2))$ to the initial eccentricity.
This ratio also remains close to unity.
We thus conclude that none of these observables may be a good probe of thermalization.

Although our model is too crude to reproduce the absolute magnitude of HBT radii, we expect that the above ratios are less model dependent.
Our results for $\Delta R_{os}/\Delta R_o^2$ and $\Delta R_o^2/\Delta R_s^2$
are compatible with experimental data~\cite{Gombeaud2009}, but the latter have large error bars. On the other hand, we miss the value of $\epsilon_s$, which seems to be washed out by the expansion. Hydrodynamical calculations using a soft EOS, reported a fair agreement with the mesured value of $\epsilon_s$.
It thus seems that the soft equation of state of QCD is responsible for the reduced eccentricity seen in the data.

\section*{Conclusion}

We have carried out a systematic study of how HBT observables evolve with the degree of thermalization ($K$). Our results show 
that HBT observables depend only weakly on $K$. The decrease of the radii with $p_t$ is already expected from initial conditions and
is only slightly enhanced by collective flow. We have also shown that for realistic values of $K$, 
inferred from elliptic flow study, the ratio $R_o/R_s$ is lower than $1.2$. Partial 
thermalization solves most of the HBT puzzle.


\section*{Acknowledgments}
T.L. is supported by the Academy of Finland, contract 126604.

\end{document}